\documentclass[twocolumn,aps,prb,showpacs,amsmath,amssymb]{revtex4}

\usepackage{graphicx}% Include figure files
\usepackage{dcolumn}% Align table columns on decimal point
\usepackage{bm}% bold math

\newcommand{\ep}{\varepsilon}

\begin{document}

\preprint{}

\title{Revised theory of the magnetic surface anisotropy of impurities in
metallic mesoscopic samples}

\author{O. \'Ujs\'aghy$^a$, L. Szunyogh$^{b}$, A. Zawadowski$^{a,b,c}$}
\affiliation{$^{a}$Budapest University of Technology and Economics, Institute
  of Physics and Research group Physics of Condensed Matter of Hungarian
  Academy of Sciences H-1521 Budapest, Hungary} 
\affiliation{$^{b}$Department of Theoretical Physics, Budapest University of
  Technology and Economics, H-1521 Budapest, Hungary}
\affiliation{$^{c}$Research Institute for Solid State Physics, POB 49, H-1525
Budapest, Hungary}

\date{\today}

\begin{abstract}
  In several experiments the magnitude of the contribution of magnetic
  impurities to the Kondo resistivity shows size dependence in mesoscopic
  samples. It was suggested ten years ago that magnetic surface anisotropy can
  be responsible for the size dependence in cases where there is strong
  spin-orbit interaction in the metallic host. The anisotropy energy has the
  form $\Delta E=K_d ({\bf n}{\bf S})^2$ where ${\bf n}$ is the vector
  perpendicular to the plane surface, ${\bf S}$ is the spin of the magnetic
  impurity and $K_d>0$ is inversely proportional to distance $d$ measured from
  the surface. It has been realized that in the tedious calculation an
  unjustified approximation was applied for the hybridizations of the host
  atom orbitals with the conduction electrons which depend on the position of
  the host atoms. Namely, the momenta of the electrons were replaced by the
  Fermi momentum $k_F$.  That is reinvestigated considering the $k$-dependence
  which leads to singular energy integrals and in contrary to the previous
  result $K_d$ is oscillating like $\sin (2 k_F d)$ and the distance
  dependence goes like $1/d^3$ in the asymptotic region. As the anisotropy is
  oscillating, for integer spin the ground state is either a singlet or a
  doublet depending on distance $d$, but in the case of the doublet there is
  no direct electron induced transition between those two states at zero
  temperature.  Furthermore, for half-integer ($S > 1/2$) spin it is always a
  doublet with direct transition only in half of the cases.
\end{abstract}


\maketitle 

\section{Introduction}

\begin{figure}[htbp]
  \centering
  \includegraphics[height=2cm]{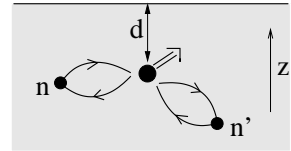}
  \caption{The magnetic impurity at a distance $d$ from the surface in a
    metallic host with homogeneously dispersed spin-orbit scatterers labeled
    by $n$.}
  \label{fig1}
\end{figure}

There are substantial experimental evidences that the amplitude of the Kondo
effect due to magnetic impurities in metallic samples of limited size are
reduced \cite{experiments,UZ_Japan} but with unchanged Kondo temperature. That
indicates that not all of the impurities contribute in the same way. There
were early speculations that this reduction appears where the sample size is
comparable with the Kondo screening cloud. This is incorrect as the Kondo
coupling is local and the only relevant energy scale to be compared with the
Kondo temperature is the level spacing of the conduction electrons which is
e.g. zero for semi-infinite samples. Later it was suggested \cite{UZ} that a
magnetic surface anisotropy can develop due to the spin-orbit interaction in
the host metal, which has the form
\begin{equation}
  \label{Hamilton}
  H=K_d (S_z)^2
\end{equation}
where the constant $K_d$ depends on the distance measured from the surface of
the sample and $S_z$ is the component of the impurity spin perpendicular to
the surface (see Fig.~\ref{fig1}). In those papers \cite{UZ_Japan,UZ,UZ1} it
was stated that surprisingly $K_d$ for large distances is always positive and
decays with the first power of the distance. That result was not questioned in
Ref.~[\onlinecite{belgak}]. Recently, one of the authors (L.Sz.) has called
the attention to an unjustified approximation in the previous lengthy
calculation \cite{UZ,UZ1} which can be responsible for the very surprising
results. That approximation was that in the hybridizations of the host atom
orbitals with the conduction electrons which depend on the position of the
host atoms (see Eq.(3) and (9) of Ref.~[\onlinecite{UZ1}]), the momenta of the
electrons were replaced by the Fermi momentum $k_F$.  That is even not the
case in the derivation of the Friedel oscillation \cite{Kittel}.

Meanwhile great efforts has been made to derive the surface anisotropy by
using electronic structure calculations. First Szunyogh and Gyorffy
calculated the anisotropy in semi-infinite $Au$ host for $Fe$ impurities
\cite{SZGY}. They found that $K_d$ is an oscillating function of the distance
$d$ and the amplitude falls as $1/d^2$. That was a calculation of mean field
type and the discrepancy between those and the analytical ones was not
surprising as in the latter the diagrams calculated are beyond the mean field
approximation. Recently, Szunyogh, Zar\'and, Gallego, $\mathrm{Mu\tilde{n}oz}$ 
and Gyorffy \cite{SZ2006} have developed another model,
where the spin-orbit interaction was placed on the $d$-level of the impurity
instead of the host. They considered the Friedel oscillation in the density of
states nearby the Fermi energy due to the presence of the surface and the
different $d$-orbitals of the impurity coupled differently to these
oscillations and that is realized in the oscillating anisotropy decaying as
$1/d^2$. It is interesting to note that the Hartree-Fock mean field
approximation and the diagram beyond that play equal role. We are also informed
that very elaborate calculations by A. Szilva, L. Szunyogh, G. Zar\'and, and
M.C. Mu\~noz \cite{SZprogress} are in progress where the spin-orbit
interaction in the host is considered.

The relative importance of the spin-orbit interactions on the d-level and the
host material must be very specific for which impurity atom and host metal are
considered and the final answer can be given only by detailed electronic
structure calculations.

The present analytical calculation is focused at the oscillating behavior and
the decay rate of the $K_d$ function. All of the results are obtained in the
large distance asymptotic region, as the preasymptotic calculation would be
even more difficult. The consequences of the oscillating
behavior will be discussed at the end of the paper. The main goal is to
present correct analytical result to be compared in the future with the
numerical results which may lead to the resolution of the present
discrepancies. The comparison with the experiments is left for the future when
the numerical calculation will be completed by which the very relevant
preasymptotic behavior is also achievable.

The Paper is organized as follows. In Section \ref{sec:2} we present the
outline of the problem and call the attention to the differences in the
calculation compared to the earlier works \cite{UZ,UZ1}. 
In Section \ref{sec:3} the integrals with respect to the energies are performed
which are crucial in obtaining the correct form of the anisotropy. The
consequences are analyzed in the Conclusions. Some of the matrix elements and
further details of the calculations are presented in Appendices~\ref{app1} and
\ref{app2}, respectively.

\section{The outline of the problem}
\label{sec:2}

The magnetic impurity scatters the electron in the $l=2$ orbital channel and
the spin-orbit scattering is also restricted to those \cite{UZ,UZ1}. As in
Ref.~[\onlinecite{UZ,UZ1}] we start with the conduction electron propagator
leaving and arriving at the impurity and in meantime it is scattered by one of
the heavy host atom due to strong spin-orbit scattering. The Green's function
has a simple form in the coordinate system where the impurity is in the origin
and the scattering atom is on the $z$-axis at a position $R_n$, which is
called the local system labeled by $n$. The Anderson model \cite{Anderson} is
used for the scattering $d$-levels of the host atom and the spin-orbit
scattering is assumed to happen on the $d$-level and that determines the
symmetry. Following Ref.~[\onlinecite{UZ1}] the conduction electron Green's
function in first order of the spin-orbit coupling is
\begin{eqnarray}
G_{km\sigma,k'm'\sigma'}(i\omega_n)&=&
\frac{\delta_{kk'}\delta_{mm'}
\delta_{\sigma\sigma'}}{i\omega_n-\varepsilon_k}\nonumber \\ 
&&\hspace{-3cm}+\sum\limits_{n}
\frac{1}{i\omega_n-\varepsilon_k} \frac{W_{\substack{kk'\\
\substack{mm'\\\sigma\sigma'}}}(R_n)}{(i\omega_n-\omega_d)(i\omega_n-\omega_d)}
\frac{1}{i\omega_n-\varepsilon_{k'}},
\label{elprop}
\end{eqnarray}
where now the $k$-dependence in $W$ and the $\omega$-dependence in the
$d$-level Green function is kept i.e.
$G_d(i\omega_n)=\frac{1}{i\omega_n-\omega_d}$ where $\omega_d=\ep_d-i\Delta$
and $\ep_d$ (measured from the Fermi level) and $\Delta$ are the energy and
width of the $d$-level, respectively and $-2\leq m\leq 2$ for the conduction
electrons.

In Eq.~(\ref{elprop}) now
\begin{eqnarray}
&&\hspace{-0.5cm}W_{\substack{kk'\\\substack{mm'\\\sigma\sigma'}}}(R_n)=
\nonumber \\
&&\hspace{-0.5cm}\lambda V^2
\biggl(B^+(k,k')\sigma^-+B^-(k,k')\sigma^++B^z(k,k')
\sigma^z\biggr)_{mm'\sigma\sigma'}.
\label{W}
\end{eqnarray}
which follows from a similar calculation like in Ref.~[\onlinecite{UZ1}] and
$\lambda$ is the strength of the spin-orbit interaction. $B^\pm(k,k')$ and
$B^z(k,k')$ are 5$\times$5 matrices in the quantum number $m$, having the form
\begin{subequations}
\begin{equation}
B^+_{mm'}(k,k')=\sqrt{(3+m')(2-m')} v_{m}(k) v_{m'}(k') \delta_{m,m'+1},
\end{equation}
\begin{equation}
B^-_{mm'}(k,k')=\sqrt{(3-m')(2+m')} v_{m}(k) v_{m'}(k') \delta_{m,m'-1},
\end{equation}
\begin{equation}
B^z_{mm'(k,k')}=m v_{m}(k) v_{m'}(k') \delta_{m,m'},
\end{equation}
\end{subequations}
where the $v_m(k)$ matrix elements given in Appendix~\ref{app1} are the same
as in Eq.~(13) of Ref.~[\onlinecite{UZ1}]. These are combinations of
oscillating functions like $\sin(k R_n)$ and $\cos(k R_n)$ combined with
powers like $(k R_n)^{-m-n}$ ($n=1,2,\dots$).

\begin{figure}[htbp]
  \centering
  \includegraphics[height=2cm]{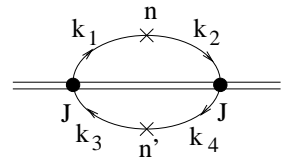}
  \caption{The self-energy diagram for the impurity spin. The
double line represents the spin, the single one the conduction
electrons. The solid circles stand for the exchange interaction and the
$\times$ labeled by $n$ for the effective spin-orbit interaction on
the orbital of the host atom at ${\bf R}_n$.}
  \label{fig2}
\end{figure}
The next step of the calculation is the rotation of the coordinate system from
the $n$-local one to that one where the $z$-axis is perpendicular to the
surface. The angle between the $z$-axis of the old ($z_n$) and the new ($z$)
coordinate system is labeled by $\Theta_n$. The calculation of
the spin factor of the self-energy diagram (see Fig.~\ref{fig2}) giving the
anisotropy for the impurity spin is also similar to the original one
(see Eqs.(21)-(25) of Ref.~[\onlinecite{UZ1}]).

The average over the positions of the scattering atoms $R_n$ and $R_{n'}$ must
be performed for the whole volume of the sample, separately. For the sake of
simplicity the continuous limit is applied outside the impurity spin. As it
was shown in the earlier works \cite{UZ,UZ1,UZ_Japan} in order to get the
dominant contribution one of $n$-s is nearby the impurity and the other one
experiences the existence of the surface at large distances.

The analytical part of the self-energy diagram Fig.~\ref{fig2} now is,
however, more complicated as $W$-s depend also on four different electronic
momenta and the corresponding energies appear in the energy denominators of
the electron Green's functions. In this way the prefactors also depend on the
momenta and that plays a crucial role in the following.  

For the sake of simplicity we consider the conduction electron band with
constant density of states $\rho_0$ in the energy interval $-D<\ep<D$ where
$\ep$ is measured from the Fermi energy and we will assume linear dispersion,
i.e. the corresponding $k$-values are $k=k_F+\frac{\ep}{v_F}$ where $v_F$ is
the Fermi velocity.

As in a noble metal host like $Cu$, $Ag$ or $Au$ the $d$-band is below the
Fermi energy, it does not give a new singularity in the energy integrals (see
Section~\ref{sec:3}), thus we can replace the $d$-level propagator by a
constant $\ep_0^{-1}$. 

Calculating the contribution of the diagram in Fig.~\ref{fig2} we applied the
Abrikosov's pseudofermion technique \cite{Abrikosov} for the spin.  After
performing the summation over the Matsubara-frequencies we get
\begin{eqnarray}
&&\hspace{-0.8cm}\rho_0^4\int\limits_{-D}^{D}d\ep_1
\int\limits_{-D}^{D}d\ep_2
\int\limits_{-D}^{D}d\ep_3\int\limits_{-D}^{D}d\ep_4\nonumber \\
&&\hspace{-0.8cm}\biggl \{
\frac{1}{\ep_1-\ep_2}\frac{(1-n_F(\ep_1))n_F(\ep_3)}
{\ep_3+\omega-\ep_1}\frac{1}{\ep_3-\ep_4}\nonumber \\
&&\hspace{-0.6cm}+(\ep_1\leftrightarrow\ep_2)+(\ep_3\leftrightarrow\ep_4)+
(\ep_1\leftrightarrow\ep_2\mbox{ and }\ep_3\leftrightarrow\ep_4)\biggr\}
\nonumber \\
&&\hspace{-0.8cm}\times \biggl \{S^2
F_1(R_n,\theta_n,R_{n'},\theta_{n'};k_1,k_2,k_3,k_4)\nonumber \\
&&\hspace{-0.2cm}+S_z^2 F_2(R_n,\theta_n,R_{n'},\theta_{n'};k_1,k_2,k_3,k_4)
\biggr \},
\label{jarulek}
\end{eqnarray}
where $\omega$ is the energy of the spin after analytical continuation and the
$F_1$, $F_2$ functions given in Appendix~\ref{app1} are defined in the same
way as in Ref.~[\onlinecite{UZ1}].

As that diagram contains two host atoms, averages have to be taken over
$n$ and $n'$. According to
our simple model \cite{UZ1} the anisotropy factor \cite{UZ,UZ1} follows as
\begin{eqnarray}
&&\hspace{-1.3cm}K=\frac{1}{a^6}\int d^3R_n\int d^3R_{n'}\rho_0^4
\int\limits_{-D}^{D}d\ep_1\int\limits_{-D}^{D}d\ep_2
\int\limits_{-D}^{D}d\ep_3\int\limits_{-D}^{D}d\ep_4\nonumber \\
&&\hspace{-0.8cm}\biggl\{\frac{1}{\ep_1-\ep_2}\frac{(1-n_F(\ep_1))n_F(\ep_3)}
{\ep_3+\omega-\ep_1}\frac{1}{\ep_3-\ep_4}\nonumber \\
&&\hspace{-0.6cm}+(\ep_1\leftrightarrow\ep_2)+(\ep_3\leftrightarrow\ep_4)+
(\ep_1\leftrightarrow\ep_2\mbox{ and }\ep_3\leftrightarrow\ep_4)\biggr\}
\nonumber \\
&&\hspace{+0.5cm}\times F_2(R_n,\theta_n,R_{n'},\theta_{n'};k_1,k_2,k_3,k_4),
\end{eqnarray}
where $a^3$ is the size of the volume per host atom. 

Changing the order of the summation over the host atoms with the energy 
integrals, the former can be evaluated in a similar way like in 
Ref.~[\onlinecite{UZ1}]. 

Eq.(29) of Ref.~[\onlinecite{UZ1}] now reads
\begin{eqnarray}\label{Kth}
&&\hspace{-0.8cm}K=\rho_0^4\int\limits_{-D}^{D}d\ep_1\int
\limits_{-D}^{D}d\ep_2
\int\limits_{-D}^{D}d\ep_3\int\limits_{-D}^{D}d\ep_4\nonumber \\
&&\hspace{-0.5cm}\biggl\{\frac{1}{\ep_1-\ep_2}\frac{(1-n_F(\ep_1))n_F(\ep_3)}
{\ep_3+\omega-\ep_1}\frac{1}{\ep_3-\ep_4}\nonumber \\
&&\hspace{-0.6cm}+(\ep_1\leftrightarrow\ep_2)+(\ep_3\leftrightarrow\ep_4)+
(\ep_1\leftrightarrow\ep_2\mbox{ and }\ep_3\leftrightarrow\ep_4)
\biggr\}\nonumber \\
&&\hspace{-0.8cm}\times \biggl \{
\frac{1}{a^6}\int\limits_d^{\infty} dR_n R_n^2\int\limits_{r_0}^d dR_{n'}
R_{n'}^2 \bigl [J_1(R_n,R_{n'};k_1,k_2,k_3,k_4)\nonumber \\
&&\hspace{+2.5cm}+J_1(R_n,R_{n'};k_3,k_4,k_1,k_2)\bigr ]\nonumber \\
&&\hspace{-0.6cm}+\frac{1}{a^6}\int\limits_d^{\infty} dR_n R_n^2
\int\limits_d^{\infty} dR_{n'}
R_{n'}^2 J_2(R_n,R_{n'};k_1,k_2,k_3,k_4)\nonumber \\
&&\hspace{-0.4cm}\biggr \},
\end{eqnarray}
where $r_0$ is a short distance cutoff in range of the atomic radius, and
\begin{eqnarray}
J_1(R_n,R_{n'};k_1,k_2,k_3,k_4)&=&(2\pi)^2\int\limits_{\theta_{n,min}}^{\pi}
d\theta_n\sin\theta_n\nonumber \\
&&\hspace{-4.5cm}\times\int\limits_0^{\pi}d\theta_{n'}\sin\theta_{n'} 
F_2(R_n,R_{n'},\theta_n,\theta_{n'};k_1,k_2,k_3,k_4),
\label{J1}\end{eqnarray}
\begin{eqnarray}
J_2(R_n,R_{n'};k_1,k_2,k_3,k_4)&=&(2\pi)^2\int\limits_{\theta_{n,min}}^{\pi}
d\theta_n\sin\theta_n\nonumber \\
&&\hspace{-5cm}\times\int\limits_{\theta_{n',min}}^{\pi}d\theta_{n'}
\sin\theta_{n'}
F_2(R_n,R_{n'},\theta_n,\theta_{n'};k_1,k_2,k_3,k_4),
\label{J2}
\end{eqnarray}
where $\theta_{n,min}=arccos(d/R_n)$, $\theta_{n',min}=arccos(d/R_{n'})$

Since according to the earlier works \cite{UZ,UZ1,UZ_Japan} the largest
contribution comes from the first part of Eq.~(\ref{Kth}) corresponding to
$J_1$ we will consider that.

The evaluation of the integrals with respect to $\theta_n$
and $\theta_{n'}$ gives
\begin{eqnarray}
  \label{J1out}  
&&\hspace{-1cm}
J_1(R_n,R_{n'};k_1,k_2,k_3,k_4)=\frac{4}{15} J^2\bigl (\frac{2\pi V^2\lambda}
     {\varepsilon_0^2}\bigr )^2 \frac{d (R_n^2-d^2)}{R_n^3}\nonumber \\
     &&\hspace{-1cm}\times\bigl [3 v_0(k_2,R_n) v_1(k_1,R_n) + 
      3 v_0(k_1,R_n) v_1(k_2,R_n) \nonumber \\
     &&\hspace{-1cm} - 2 v_1(k_1,R_n) v_1(k_2,R_n) + 
      2 v_1(k_2,R_n) v_2(k_1,R_n) \nonumber \\
     &&\hspace{-1cm}+ 2 v_1(k_1,R_n) v_2(k_2,R_n) - 
      8 v_2(k_1,R_n) v_2(k_2,R_n)\bigr ]\nonumber \\
     &&\hspace{-1cm}\times\bigl [3 v_0(k_4,R_{n'}) v_1(k_3,R_{n'}) + 
      3 v_0(k_3,R_{n'}) v_1(k_4,R_{n'}) \nonumber \\
     &&\hspace{-1cm}+ v_1(k_3,R_{n'}) v_1(k_4,R_{n'}) + 
      2 v_1(k_4,R_{n'}) v_2(k_3,R_{n'})\nonumber \\
     &&\hspace{-1cm}+ 2 v_1(k_3,R_{n'}) v_2(k_4,R_{n'}) + 
      4 v_2(k_3,R_{n'}) v_2(k_4,R_{n'})\bigr ].
\end{eqnarray}
 
If $k_1$, $k_2$, $k_3$, $k_4$ are replaced by $k_F$ that gives back the half
of Eq.~(B2) of Ref.~[\onlinecite{UZ1}].

After a straightforward calculation of the integrals with respect to $R_n$ and
$R_{n'}$ (see Appendix~\ref{app2}) the first part of Eq.~(\ref{Kth})
corresponding to $J_1$ reads
\begin{eqnarray}\label{Ksim}
&&\hspace{-0.8cm}\rho_0^4\int\limits_{-D}^{D}d\ep_1
\int\limits_{-D}^{D}d\ep_2
\int\limits_{-D}^{D}d\ep_3
\int\limits_{-D}^{D}d\ep_4\nonumber \\
&&\hspace{-1cm}\biggl\{\frac{1}{\ep_1-\ep_2}\frac{(1-n_F(\ep_1))n_F(\ep_3)}
{\ep_3+\omega-\ep_1}\frac{1}{\ep_3-\ep_4}\nonumber \\
&&\hspace{-0.6cm}+(\ep_1\leftrightarrow\ep_2)+(\ep_3\leftrightarrow\ep_4)+
(\ep_1\leftrightarrow\ep_2\mbox{ and }\ep_3\leftrightarrow\ep_4)
\biggr\}\nonumber \\
&&\hspace{-1cm}\times\frac{4}{15} J^2\bigl (\frac{2\pi V^2\lambda}{\ep_0^2}
\bigr )^2\nonumber \\
&&\hspace{-1.1cm}\times\bigl [C(k_1 d,k_2 d) D(k_3 d,k_4 d)
+C(k_3 d,k_4 d) D(k_1 d,k_2 d)\bigr ],
\end{eqnarray}
where the functions $C$ and $D$ are given by Eqs.~(\ref{Cujabb}) and
(\ref{Dujabb}), respectively. 

As $C$ and $D$ are symmetric in their variables
we can change the integration variables according to the changes indicated 
in the energy dependent factor in the integrand of Eq.~(\ref{Ksim}) resulting
in a simpler form like
\begin{eqnarray}\label{Ksimveg}
&&\hspace{-0.8cm}\rho_0^4\int\limits_{0}^{D}d\ep_1
\int\limits_{-D}^{D}d\ep_2
\int\limits_{-D}^{0}d\ep_3
\int\limits_{-D}^{D}d\ep_4\nonumber \\
&&\hspace{-1cm}\biggl\{\frac{1}{\ep_1-\ep_2}\frac{1}
{\ep_3+\omega-\ep_1}\frac{1}{\ep_3-\ep_4}\biggr\}
\,4\frac{4}{15} J^2\bigl (\frac{2\pi V^2\lambda}{\ep_0^2}
\bigr )^2\nonumber \\
&&\hspace{-1.1cm}\times\bigl [C(k_1 d,k_2 d) D(k_3 d,k_4 d)
+C(k_3 d,k_4 d) D(k_1 d,k_2 d)\bigr ],
\end{eqnarray}
where we have exploited the $1-n_F(\ep_1)$ and $n_F(\ep_3)$ factors as
well.

\section{The energy integrals}
\label{sec:3}

In the following the asymptotic behavior for large distances $d$ is
considered, therefore, only the leading order in $\frac{1}{d}$ is
kept everywhere. 

For large distances the radial electronic wave functions are fast oscillating
as the energy is changed. These fast oscillations lead to essential
cancellations. In order to keep track of the cancellations in the limit
$d\rightarrow\infty$, the Riemann theorem with the first asymptotic correction
is applied in the following form \cite{Riemann}
\begin{subequations}\label{Riemann}
\begin{equation}
\int\limits_{a}^{b} d s f(s) \cos (x s)\sim 
\frac{f(b)\sin (x b)}{x}-\frac{f(a)\sin (x a)}{x}
\end{equation}
and
\begin{equation}
\int\limits_{a}^{b} d s f(s) \sin (x s)\sim 
\frac{f(a)\cos (x a)}{x}-\frac{f(b)\cos (x b)}{x}
\end{equation}
\end{subequations}
which is valid in the leading order in $1/x$ where $f$ must be integrable. 

Let us consider the integrations with respect to the energies.  In the
first part of Eq.~(\ref{Ksimveg}) the integral with respect to $\ep_2$ is
\begin{equation}\label{ep2}
\int\limits_{-D}^{D}d\ep_2 \frac{C(k_1 d,k_2 d)}{\ep_1-\ep_2}
=\int\limits_{k_F-\frac{D}{v_F}}^{k_F+\frac{D}{v_F}} dk_2
\frac{C(k_1 d,k_2 d)}{k_1-k_2}.
\end{equation}

Introducing a new integration variable $\Delta k=k_2-k_1$ and using
linear dispersion $\ep=v_F (k-k_F)$, the integral reads
\begin{eqnarray}\label{ep2uj}
&&\hspace{-0.7cm}-\int\limits_{-k_1+k_F
-\frac{D}{v_F}}^{-k_1+k_F+\frac{D}{v_F}} 
d(\Delta k)\frac{C(k_1 d,(k_1+\Delta k) d)}{\Delta k}\nonumber \\
&&\hspace{-0.7cm}=-\frac{d^3}{a^3}
\int\limits_{-k_1+k_F-\frac{D}{v_F}}^{-k_1+k_F+\frac{D}{v_F}}
d(\Delta k)\nonumber \\
&&\times\bigl [
f_{c+}(k_1 d,(k_1+\Delta k) d) (\cos[2 k_1 d] \frac{\cos[\Delta k d]}{\Delta
  k}\nonumber \\
&&\hspace{+3.5cm}-\sin[2 k_1 d] \frac{\sin[\Delta k d]}{\Delta
  k})\nonumber \\
&&+f_{c-}(k_1 d,(k_1+\Delta k) d) \frac{\cos[\Delta k d]}{\Delta
  k}\nonumber \\
&&+f_{s+}(k_1 d,(k_1+\Delta k) d) (\sin[2 k_1 d] \frac{\cos[\Delta k d]}{\Delta
  k}\nonumber \\
&&\hspace{+3.5cm}+\cos[2 k_1 d] \frac{\sin[\Delta k d]}{\Delta
  k})\nonumber \\
&&+d f_{s-}(k_1 d,(k_1+\Delta k) d) \sin[\Delta k d]\nonumber \\
&&+\Delta k d^2 f_{ci}(k_1 d,(k_1+\Delta k) d)\nonumber \\
&&\hspace{+0.5cm}\times\bigl (Ci[(2 k_1+\Delta k) d]
-Ci[\Delta k d]\bigr )\bigr ],
\end{eqnarray}
where we used the symmetry property of the cosine integral function
$Ci[-x]=Ci[x]$ and trigonometrical identities\cite{Abramowitz}.

To evaluate the terms containing the $Ci[x]$ function we use 
\begin{equation}\label{Ciatir}
Ci[x]=-\int\limits_x^{\infty} du\frac{\cos u}{u}=
-\int\limits_1^{\infty} dv\frac{\cos x v}{v}
\end{equation}
and change the order of the integrations with respect to $v$ and
$\Delta k$.

Due to the cosine and sine functions in the integrand, the integral is
determined by the singularity at $\Delta k=0$ ($k_2=k_1$). Searching
for that we expand the
$f$-functions around $k_1 d$ in their second variables and then drop
the terms which are not singular at
$\Delta k=0$. Then the integral Eq.~(\ref{ep2uj}) is
\begin{eqnarray}\label{ep2meguj}
&&\hspace{-0.5cm}-\frac{d^3}{a^3}
\int\limits_{-k_1+k_F-\frac{D}{v_F}}^{-k_1+k_F+\frac{D}{v_F}}
d(\Delta k) \nonumber \\
&&\biggl \{\frac{\cos[\Delta k d]}{\Delta
  k}\bigl (f_{c+}(k_1 d,k_1 d)\cos[2 k_1 d]\nonumber \\
&&\hspace{0.8cm}+f_{c-}(k_1 d,k_1 d)
+f_{s+}(k_1 d,k_1 d)\sin[2 k_1 d]\bigr )\nonumber \\
&&+\frac{\sin[\Delta k d]}{\Delta k}\bigl (
-f_{c+}(k_1 d,k_1 d)\sin[2 k_1 d]\nonumber \\
&&\hspace{1.2cm}+f_{s+}(k_1 d,k_1 d)\cos[2 k_1 d]
\bigr )
\biggr\}.
\end{eqnarray}

The range of the integrations can be extended to
$-\infty\rightarrow \infty$ as those integrals are independent of $d$, while
the added parts are fast oscillating and, therefore, they are $\mathcal{O}
(\frac{1}{k_F d})$ as it can be proved by using the Riemann theorem given by
Eq.~(\ref{Riemann}) also. Then using
\begin{equation}
\int\limits_{-\infty}^{\infty} 
d(\Delta k) \frac{\cos[\Delta k d]}{\Delta k}=0
\end{equation}
and
\begin{equation}
\int\limits_{-\infty}^{\infty} 
d(\Delta k) \frac{\sin[\Delta k d]}{\Delta k}=\pi,
\end{equation}
we get for Eq.~(\ref{ep2})
\begin{eqnarray}
  \label{ep2veg}
&&\hspace{-0.5cm}\int\limits_{-D}^{D}d\ep_2 
\frac{C(k_1 d,k_2 d)}{\ep_1-\ep_2}=\nonumber \\
&&\hspace{-0.3cm}=\frac{d^3}{a^3}\pi\biggl [f_{c+}(k_1 d,k_1 d) \sin[2 k_1
 d]-f_{s+}(k_1 d,k_1 d) \cos[2 k_1 d]\biggr ]\nonumber \\
&&\hspace{-0.3cm}\approx \frac{d^3}{a^3}\pi\biggl [
\frac{3825}{4 (k_1 d)^6}\sin[2 k_1
 d]-\frac{225}{2 (k_1 d)^5} \cos[2 k_1 d]\biggr ]\nonumber \\
&&\hspace{-0.3cm}\approx -\frac{d^3}{a^3}\frac{225\pi}{2 (k_1 d)^5} 
\cos[2 k_1 d],
\end{eqnarray}
where we kept only the leading order contribution in $1/k_1 d$ as
$k_1 d\geq k_F d\gg 1$ according to the range of the integration with
respect to $k_1$ ($\ep_1$) in Eq.~(\ref{Ksimveg}).

Let's turn to the integration with respect to $\ep_1$ in the first
part of Eq.~(\ref{Ksimveg}) i.e. to
\begin{eqnarray}
  \label{ep1sima}
&&\hspace{-1.5cm}\int\limits_{0}^{D}\frac{d\ep_1}{\ep_3+\omega-\ep_1}
\int\limits_{-D}^{D}d\ep_2 
\frac{C(k_1 d,k_2 d)}{\ep_1-\ep_2}=\nonumber \\
&&\hspace{-1.5cm}=-\frac{d^3}{a^3}\int\limits_{0}^{D}
\frac{d\ep_1}{\ep_3+\omega-\ep_1}
\frac{225\pi}{(k_1 d)^5}\frac{1}{2}\cos[2 k_1 d]\nonumber \\
&&\hspace{-1.5cm}=-\frac{d^3}{a^3}
\int\limits_{k_F}^{k_F+\frac{D}{v_F}}
\frac{d k_1}{k_3+\frac{\omega}{v_F}-k_1}
\frac{225\pi}{(k_1 d)^5}\frac{1}{2}\cos[2 k_1 d].
\end{eqnarray}

As $\ep_3<0$ ($k_3<k_F$) in Eq.~(\ref{Ksimveg}) and $\omega\approx 0$ the
integrand has no singularities in the range of the integration, thus
in order to find the leading order contribution in $1/k_F d$ we can
apply the Riemann theorem given by Eq.~(\ref{Riemann}).
Then Eq.~(\ref{ep1sima}) is
\begin{eqnarray}\label{ep1vegneg}
&&\hspace{-1.5cm}\int\limits_{0}^{D}\frac{d\ep_1}{\ep_3+\omega-\ep_1}
\int\limits_{-D}^{D}d\ep_2 
\frac{C(k_1 d,k_2 d)}{\ep_1-\ep_2}\approx\nonumber \\
&&\hspace{-1cm}\approx\frac{d^3}{a^3}
\frac{v_F}{\ep_3+\omega}\frac{225\pi}{(k_F d)^5}\frac{1}{2}
\frac{\sin[2 k_F d]}{2 d}
\nonumber \\
&&\hspace{-0.5cm}=\frac{225\pi}{(k_F a)^3}
\frac{\ep_F}{\ep_3+\omega}\frac{1}{4}
\frac{\sin[2 k_F d]}{(k_F d)^3}
\end{eqnarray}
in leading order in $1/k_F d$, 
where we kept only the contribution of the lower limit, as the contribution of
the upper limit of the integral in Eq.~(\ref{ep1vegneg}) is proportional to
$1/D$, thus it is of lower order.

Now we have to evaluate the remaining integrals with respect to $\ep_3$ and
$\ep_4$ in the first part of Eq.~(\ref{Ksimveg}) i.e.
\begin{equation}\label{ep3ep4}
\int\limits_{-D}^{0}d\ep_3 \frac{1}{\ep_3+\omega}
\int\limits_{-D}^{D}d\ep_4 \frac{D(k_3 d,k_4 d)}{\ep_3-\ep_4}.
\end{equation}
Starting with the first part of Eq.~(\ref{ep3ep4}) corresponding to the $g$
functions in $D$ (see Eq.~(\ref{Dujabb})), we 
introduce again new integration variable $\Delta k=k_4-k_3$ and use
linear dispersion. Thus Eq.~(\ref{ep3ep4}) reads
\begin{eqnarray}\label{ep3ep4uj}
&&\int\limits_{k_F-\frac{D}{v_F}}^{k_F} d k_3
\frac{1}{k_3-k_F+\frac{\omega}{v_F}}
\int\limits_{k_F-\frac{D}{v_F}}^{k_F+\frac{D}{v_F}} 
d k_4\frac{D(k_3 d,k_4 d)}{k_3-k_4}\nonumber \\
&&=-\int\limits_{k_F-\frac{D}{v_F}}^{k_F} d k_3
\frac{1}{k_3-k_F+\frac{\omega}{v_F}}\nonumber \\
&&\hspace{0.5cm}\times\int\limits_{-k_3+k_F-\frac{D}{v_F}}^
{-k_3+k_F+\frac{D}{v_F}} d(\Delta k)\frac{D(k_3 d,(k_3+\Delta k) d)}{\Delta k}
\end{eqnarray}
and in the first part containing the $g$
functions (see Eq.~(\ref{Dujabb})) we can repeat the
considerations used in performing the integrals with respect to $\ep_1$ and
$\ep_2$ giving
\begin{eqnarray}\label{ep3ep4ujuj}
&&\frac{d^3}{a^3}\pi\int\limits_{k_F-\frac{D}{v_F}}^{k_F} d k_3
\frac{1}{k_3-k_F+\frac{\omega}{v_F}}\nonumber \\
&&\hspace{-1cm}\times\biggl [g_{c+}(k_3 d,k_3 d) \sin[2 k_3
 d]-g_{s+}(k_3 d,k_3 d) \cos[2 k_3 d]\biggr ].
\end{eqnarray}
Since $k_F-\frac{D}{v_F}>0$ we can again apply the Riemann theorem given by
Eq.~(\ref{Riemann}). The terms coming from the lower limit of the integrals
are less by $1/D$ than the terms coming from the upper
limit of the integral which give
\begin{eqnarray}\label{ep3ep4ujujuj_1}
&&\hspace{-0.1cm}-\frac{d^3}{a^3}\pi\frac{v_F}{\omega}\nonumber \\
&&\times\biggl [g_{c+}(k_F d,k_F d) \frac{\cos[2 k_F
 d]}{2 d}+g_{s+}(k_F d,k_F d) \frac{\sin[2 k_F d]}{2 d}
\biggr ]\nonumber \\
&&\hspace{-0.5cm}\approx -\frac{d^3}{a^3}\pi\frac{v_F}{\omega}\nonumber \\
&&\times\biggl [\frac{225}{2 (k_F d)^4}\frac{\cos[2 k_F
 d]}{2 d}-\frac{1125}{(k_F d)^5}\frac{\sin[2 k_F d]}{2 d}
\biggr ]\nonumber \\
&&\hspace{-0.5cm}\approx -\frac{1}{(k_F a)^3}\frac{\ep_F}{\omega}
\frac{225\pi}{4 (k_F d)^2}\cos[2 k_F d],
\end{eqnarray}
where only the leading order contribution in $1/k_F d$ was kept.

Turning to the second part of Eq.~(\ref{ep3ep4}) corresponding to the $h$
functions in $D$ (see Eq.~(\ref{Dujabb})), after using the following
properties
\begin{equation}
  \label{htul}
  h(k_3 d,k_4 d,\frac{r_0}{d})=\frac{1}{(k_F d)^3} 
h(\frac{k_3}{k_F},\frac{k_4}{k_F},k_F r_0)
\end{equation}
for the $h_{c+/c-/s+}$
functions (see Eq.~(\ref{hs}))
and 
\begin{eqnarray}
  \label{htul2}
 &&\hspace{-0.5cm}(k_3-k_4) d h_{s-}(k_3 d,k_4
  d,\frac{r_0}{d})\nonumber \\
&&=(\frac{k_3}{k_F}-\frac{k_4}{k_F})
\frac{1}{(k_F d)^3} 
h_{s-}(\frac{k_3}{k_F},\frac{k_4}{k_F},k_F r_0)
\end{eqnarray}
for the $h_{s-}$ function, and introducing the $s=\frac{k_3}{k_F}$ and
$t=\frac{k_4}{k_F}$ new integration variables, we get
\begin{eqnarray}\label{Pwg}
&&\frac{1}{(k_F a)^3} \frac{\mathrm{P}_1(k_F r_0,\omega)}{15\pi},
\end{eqnarray}
where
\begin{equation}
  \label{P1}
\hspace{-0.1cm}\mathrm{P}_1(x,\omega)=15\pi
\int\limits_{1-\frac{D}{\ep_F}}^{1} d s
\frac{1}{s-1+\frac{\omega}{\ep_F}}\int\limits_{1-\frac{D}{\ep_F}}^
{1+\frac{D}{\ep_F}} d t\frac{H(s,t,x)}{s-t}
\end{equation}
and
\begin{eqnarray}\label{H}
H(s,t,x)&=&h_{c+}(s,t,x)\cos[(s+t) x]
\nonumber \\
&+&h_{c-}(s,t,x)\cos[(s-t) x]\nonumber \\
&+&h_{s+}(s,t,x)\sin[(s+t) x]\nonumber \\
&+&(s-t) h_{s-}(s,t,x)\sin[(s-t) x].
\end{eqnarray}
Thus the terms corresponding to the $h$ functions give $d$-independent
contribution, therefore Eq.~(\ref{ep3ep4}) is
Eq.~(\ref{Pwg}) in leading order in $1/k_F d$.

Combining that with Eqs.~(\ref{ep1vegneg}) and (\ref{Ksimveg}) we get
for the first part of Eq.~(\ref{Ksimveg}) in leading
order in $1/k_F d$
\begin{equation}\label{Ksimkesz1}
16 \ep_F (J\rho_0)^2\frac{\Delta^2\lambda^2}{\varepsilon_0^4} 
\frac{1}{(k_F a)^6}
\mathrm{P}_1(k_F r_0,\omega)\cdot\frac{\sin [2 k_F d]}{(k_F d)^3},
\end{equation}
where $\Delta=\pi V^2\rho_0$ is the width of the $d$-levels due to
hybridization \cite{Anderson}.

Turning to the second part of Eq.~(\ref{Ksimveg}), after changing the
integration variables as $\ep_1\leftrightarrow\ep_3$ and
$\ep_2\leftrightarrow\ep_4$ and performing similar calculation as before we
get in leading order with respect to $1/k_F d$
\begin{equation}\label{Ksimkesz2}
16 \ep_F (J\rho_0)^2\frac{\Delta^2\lambda^2}{\varepsilon_0^4} 
\frac{1}{(k_F a)^6}
\mathrm{P}_2(k_F r_0,\omega)\cdot\frac{\sin [2 k_F d]}{(k_F d)^3},
\end{equation}
where 
\begin{equation}\label{P2}
\hspace{-0.1cm}\mathrm{P}_2(x,\omega)=15\pi
\int\limits_{1}^{1+\frac{D}{\ep_F}} d s
\frac{1}{s-1-\frac{\omega}{\ep_F}}\int\limits_{1-\frac{D}{\ep_F}}^
{1+\frac{D}{\ep_F}} d t\frac{H(s,t,x)}{s-t}.
\end{equation}
  
Thus the anisotropy factor coming from the first part of
Eq.~(\ref{Kth}) corresponding to $J_1$ -- giving the leading
contribution \cite{UZ,UZ1,UZ_Japan} -- in leading order in $1/k_F
d$ is
\begin{equation}\label{Ksimkesz}
K=16 \ep_F (J\rho_0)^2
\frac{\Delta^2\lambda^2}{\varepsilon_0^4} 
\frac{1}{(k_F a)^6}\mathrm{P}(k_F r_0,\omega)\frac{\sin [2 k_F d]}{(k_F d)^3},
\end{equation}
where
\begin{equation}\label{P1mP2}
\mathrm{P}(x,\omega)=\mathrm{P}_1(x,\omega)
+\mathrm{P}_2(x,\omega)
\end{equation}
which is finite for $\omega=0$.

Evaluating $\mathrm{P}$ by numerical integration it turns out that for $x\sim
1$\cite{xnagy} the integrals with respect to $t$ and $s$ are dominated by the
$t=s$ and $s=1$ singularity, respectively. Thus
 \begin{eqnarray}\label{Pkoz}
&&\hspace{-2cm}\mathrm{P}(x,\omega=0)\approx 15\pi
\int\limits_{0}^{2} d s
\frac{H(s,s,x)}{s-1}\ln\biggl |\frac{s}{s-2}
\biggr |\nonumber \\
&&=15\pi H(1,1,x)\frac{\pi^2}{2}
\end{eqnarray}
where we used $D=\ep_F$.
In Table~\ref{tabP} we compare the results obtained by numerical
integration and by Eq.~(\ref{Pkoz}) for $x\sim 0.5-1.5$.
\begin{table}[htbp]
  \centering
 \begin{tabular}{|c||c|c|c|c|c|}
\hline
%\hline
x&0.5&0.75&1&1.25&1.5 \\
\hline
$\mathrm{P}^{\mathrm{num}}
(x,\omega=0.01)$&-135.8&-428.1&-916.5&-1552.9&-2215.9 \\
\hline
$\mathrm{P}^{\mathrm{appr}}
(x,\omega=0)$&-138.3&-438.4&-952.6&-1664.6&-2514.1 \\
\hline
%\hline
  \end{tabular}
\caption{Comparison of $\mathrm{P}^{\mathrm{num}}(x,\omega=0.01)$ obtained by 
numerical integration and $\mathrm{P}^{\mathrm{appr}}(x,\omega=0)$ obtained 
by Eq.~(\ref{Pkoz}) for $x\sim 0.5-1.5$.}
\label{tabP}
\end{table}

The result Eq.~(\ref{Ksimkesz}) for the anisotropy factor is essentially
different from the earlier one (see Eq.~(31) of Ref.~[\onlinecite{UZ1}])
obtained by an unjustified assumption which corresponds formally to the
approximation $C(k_1 d,k_2 d)\approx C(k_F d,k_F d)$ and $D(k_3 d,k_4
d)\approx D(k_F d,k_F d)$ in Eq.~(\ref{Ksimveg}). The main differences are
\begin{itemize}
\item it contains the oscillating factor $\sin [2 k_F d]$
\item the asymptotic distance dependence is $\frac{1}{(k_F d)^3}$ instead of 
$\frac{1}{k_F d}$, thus it is essentially weaker
\item instead of the $f(\frac{\omega}{D}) \mathrm{P}^{old}(k_F r_0)$ factor
  which contains the short range cutoff $r_0$ and estimated to be between
  $50-950$ for $\omega\approx 0$ and $k_F r_0\sim 0.5-1.5$, we have
  $\mathrm{P}^{new}(k_F r_0,\omega)$ which is for
  $\omega=0$ and $k_F r_0\sim 0.5-1.5$ between $-140$ and $-2500$,
%\item it does not contain the $f$-function which contains some of the energy
%  integrals and it was of order of unity. 
\end{itemize}
thus in the asymptotic region using the same parameters as in Eq.~(32) in
Ref.~[\onlinecite{UZ1}] 
\begin{equation}
\frac{0.01}{(d/\mathrm{\AA})^3}\,\mathrm{eV} < |K_d| < 
\frac{1.75}{(d/\mathrm{\AA})^3}\,\mathrm{eV}.
\label{becsles}
\end{equation}

Finally, the question can be raised how justified is the assumption of
homogeneous distribution of the spin-orbit scatterers. In order to give the
answer in the following the scatterers are considered homogeneously
distributed on sheets which are parallel to the surface and separated by a
distance $a$. According to the previous works \cite{UZ,UZ1} the pair of sheets
in equal distances from the impurity do not contribute to the anisotropy, thus
only the unpaired ones must be considered. The sheet $n$ is in the distance $n
a$ from the impurity and only sheets with $n>d/a$ are considered. As it was
discussed in Ref.~[\onlinecite{UZ,UZ1,UZ_Japan}] one of the two scatterers $n$
and $n'$ is nearby the impurity and the other one is far from it on one of the
sheets considered. Therefore, the contribution of the sheets are additive. The
contribution of sheet $n$ can be easily obtained from the present calculation
as
\begin{equation}
  \label{Kn}
  K_n=a\frac{\partial}{\partial d} {K_d}_{|_{d=n a}},
\end{equation}
where the derivative gives the contribution of an infinitely narrow
layer and the prefactor provides the correct normalization. Thus the
final result in the asymptotic region is
\begin{equation}
  \label{Kfin}
  K=\sum\limits^\infty_{n>\frac{d}{a}} K_n\approx
  \sum\limits_{n>\frac{d}{a}} 2 a k_F\frac{\cos (2 k_F n a)}{(k_F n a)^3},
\end{equation}
where the omitted prefactor is the same as in Eq.~(\ref{Ksimkesz}).

Thus the separate sheets contribute by different signs and amplitude. Due to
the fast decay by increasing $n$ only sheets of restricted numbers are
essential. Adding the contribution of the sheets with different signs and
amplitudes (likely randomly distributed) the final amplitude of the anisotropy
$|K|$ can be larger than $|K_d|$ thus the $\frac{1}{d^3}$ decay rate can be
somewhat reduced. The situation is different in a coherent case with $k_F
a=p\pi$ where $p$ is integer. For even $p$ the contributions have the same
sign and $\sum\limits_{n>\frac{d}{a}}\frac{1}{n^3}\sim\frac{a^2}{d^2}$ which
provides a slower decay rate.
%\vspace{1cm}

\section{Conclusions}

The amplitude of the anisotropy is oscillating and weaker than the one earlier
estimated \cite{UZ,UZ1}. In these changes the sharp edge in the $R$-integral
thus the existence of the surface is crucial. We used a uniform distribution
of the spin-orbit scatterers in space. Considering the question how the
results are changed in the case where continuum layers of the scatterers are
considered, it is argued that the overall behavior is not expected to change,
but only the amplitude can be influenced moderately. Furthermore, it is
assumed, that the spin-orbit scattering is point like, but a finite extension
$r_d$ (it is assumed that $k_F r_d<\pi$) smears somewhat the $\sin [2 k_F d]$
function in Eq.~(\ref{Ksimkesz}). The actual size of that can be estimated
only by electronic structure calculations and certainly will be included in
the recent work of one of the authors (L.Sz.)  and his coworkers in progress
\cite{SZprogress}.

\begin{figure}[htbp]
  \centering
  \includegraphics[height=4cm]{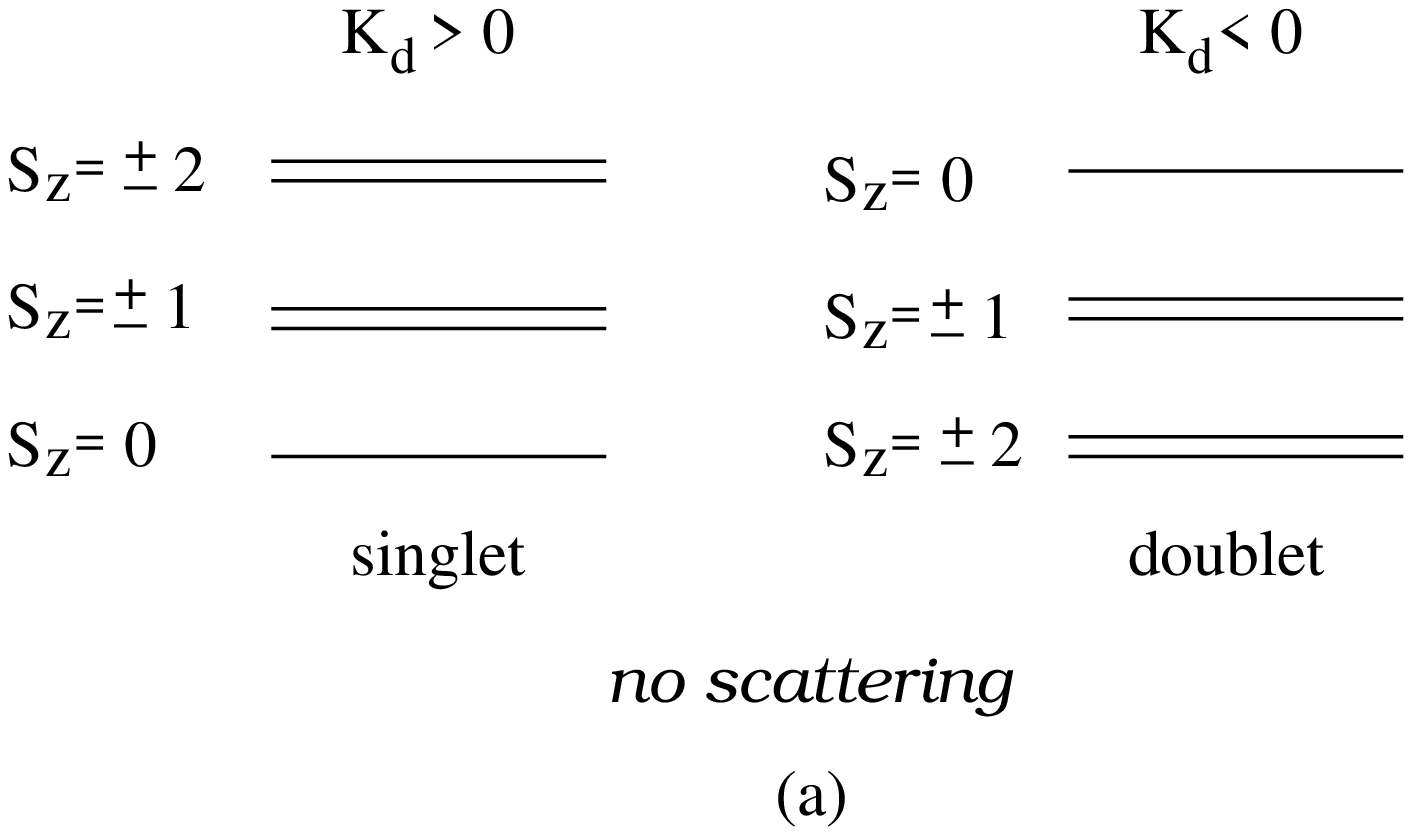}
\\[2em]
  \includegraphics[height=4cm]{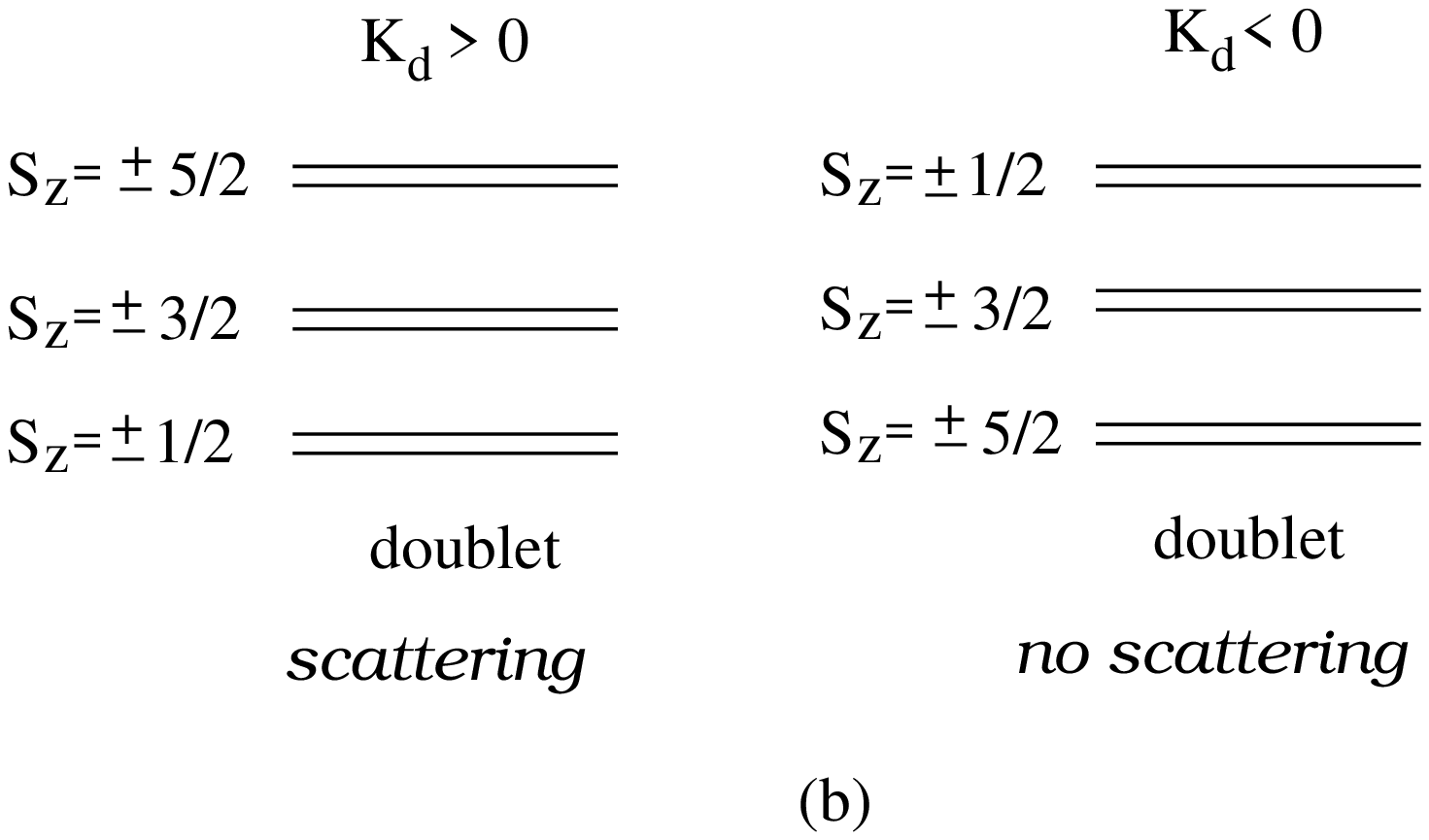}
  \caption{The level splitting due to the surface anisotropy for (a) integer 
(e.g. $S=2$) and (b) half-integer (e.g. $S=5/2$) spins. It is also
    indicated whether the electrons can be scattered by the degenerate
    ground states or not.}
  \label{fig3}
\end{figure}

The way how the spin is frozen by the surface anisotropy is
essentially different from the previous works \cite{UZ,UZ1} as $K_d$
is not always positive (see Fig.~\ref{fig3}) and also depends whether the spin
is integer or half-integer.
\begin{itemize}
\item {\it Integer spin (e.g. $S=2$):} the ground state is either singlet or
  doublet depending on the sign of $K_d$, however, the electron cannot be
  scattered by transition in the doublet as the spin momentum difference is
  $\Delta S=\pm 4$ and the turning the electron spin allows
  only $\Delta S=\pm 1$. Thus the spin at low temperature can be completely
  frozen in.
\item {\it Half-integer spin (e.g. $S=5/2$):} the ground state is always
  doublet, however, in one of the cases $\Delta S=\pm 1$ which can cause
  scattering in contrary to the case where $\Delta S=\pm 5$. Thus only half of
  the impurities can cause electron spin flips at low temperature. That is
  different from the previously assumed $K_d>0$ case.
\end{itemize}

This result shows analogies with Ref.~[\onlinecite{Romeike}] where magnetic
molecules with large spins on metallic surface were considered. The spin
levels are split in a similar way and electron induced transitions are allowed
only between certain levels.

For comparison with experiment the preasymptotic behavior is very
essential which is beyond the scope of the present paper. The
electronic calculations in progress \cite{SZprogress}
must provide those information including the amplitude
of the anisotropy as in the other model in Ref.~[\onlinecite{SZ2006}].

The present results valid in the asymptotic region can provide a good
possibility to test the numerical calculation. Detailed comparison
with experiments must wait for completing the numerical calculation
which is going to provide more necessary information.

\section*{ACKNOWLEDGMENTS}
This work was supported by Hungarian grants OTKA F043465, T046267, NF61726,
T048782, TS049881.

\appendix
\section{}
\label{app1}
The $v_m(k)$ matrix elements are the same
as in Eq.~(13) of Ref.~[\onlinecite{UZ1}]:
\begin{subequations}
\label{vkm}
\begin{eqnarray}
v_{0}(k,R_n)&=&10\biggl(\frac{\sin(k R_n)}{2 k R_n}
+\frac{3\cos(k R_n)}{(k R_n)^2}-\frac{12\sin(k R_n)}{(k R_n)^3}\nonumber \\
&-&\frac{27\cos(k R_n)}{(k R_n)^4}+\frac{27\sin(k R_n)}{(k R_n)^5}
\biggr),
\end{eqnarray}
\begin{eqnarray}
v_{1}(k,R_n)&=&15\biggl( -\frac{\cos(k R_n)}{(k R_n)^2}
+\frac{5\sin(k R_n)}{(k R_n)^3}+\frac{12\cos(k R_n)}{(k R_n)^4}\nonumber \\
&-&\frac{12\sin(k R_n)}{(k R_n)^5}\biggr),
\end{eqnarray}
\begin{equation}
v_{2}(k,R_n)=15\biggl(-\frac{\sin(k R_n)}{(k R_n)^3}
-\frac{3\cos(k R_n)}{(k R_n)^4}+\frac{3\sin(k R_n)}{(k R_n)^5}
\biggr).
\end{equation}
\end{subequations}

The $F_1$ and $F_2$ functions are defined in the same way as in
Ref.~[\onlinecite{UZ1}], namely
\begin{subequations}
\begin{equation}
F_1(R_n,\theta_n,R_{n'},\theta_{n'};k_1,k_2,k_3,k_4)=
\frac{2 J^2}{\ep_0^4} f_{1122},
\end{equation}
and
\begin{eqnarray}
&&\hspace{-1.5cm}F_2(R_n,\theta_n,R_{n'},\theta_{n'};k_1,k_2,k_3,k_4)
\nonumber \\
&&=\frac{2 J^2}{\ep_0^4} (f_{1111}-f_{1212}-f_{1122}),
\label{F2}
\end{eqnarray}
\end{subequations}
where
\begin{equation}
f_{\sigma_1\sigma_2\sigma_3\sigma_4}=\sum\limits_{mm'}
\tilde W_{\substack{k_1 k_2\\\substack{m_1 m_2\\\sigma_1\sigma_2}}}
(R_n,\theta_n)
\tilde W_{\substack{k_4 k_3\\\substack{m_2 m_1\\\sigma_4\sigma_3}}}(R_{n'},\theta_{n'}).
\end{equation}
$\tilde W$ gives the form of $W$ in the rotated coordinate system (where the
$z$-axis is perpendicular to the surface) given by
\begin{eqnarray}
&&\hspace{-0.8cm}\tilde{W}_{\substack{kk'\\\substack{mm'\\\sigma\sigma'}}}
(R_n,\theta_n)=
\delta_{m+\sigma,m'+\sigma'}
\sum\limits_{\substack{\bar m\bar m'\\\bar\sigma\bar\sigma'}}
d^{(2)}_{m\bar m}(\theta_n)d^{(1/2)}_{\sigma\bar\sigma}
(\theta_n)\nonumber \\
&&\hspace{0.8cm}
\cdot W_{\substack{kk'\\\substack{\bar m\bar m'\\\bar\sigma\bar\sigma'}}}(R_n)
d^{(2)}_{\bar m'm'}(-\theta_n)d^{(1/2)}_{\bar\sigma'\sigma'}
(-\theta_n)
\end{eqnarray}
where the Wigner-formula for rotation matrices \cite{Brink} was used.

\section{}
\label{app2}
Here we perform the integrations with respect to $R_n$ and $R_{n'}$
in the first part of Eq.~(\ref{Ksim}). Let us start with the integration with
respect to $R_n$, namely
\begin{eqnarray}
\tilde{C}(k_1,k_2)&:=&\frac{1}{a^3}\int\limits_d^{\infty} dR_n R_n^2 \frac{d
  (R_n^2-d^2)}{R_n^3}\nonumber \\
&&\hspace{-2.4cm}\times \bigl [3 v_0(k_2,R_n) v_1(k_1,R_n) + 
      3 v_0(k_1,R_n) v_1(k_2,R_n)\nonumber \\
&&\hspace{-2.3cm}- 2 v_1(k_1,R_n) v_1(k_2,R_n) + 
      2 v_1(k_2,R_n) v_2(k_1,R_n)\nonumber \\
&&\hspace{-2.3cm}+ 2 v_1(k_1,R_n) v_2(k_2,R_n) - 
      8 v_2(k_1,R_n) v_2(k_2,R_n)\bigr ].
\end{eqnarray}

After substituting the $v_m(k, R_n)$ matrix elements given by Eq.~(\ref{vkm})
and introducing the dimensionless integration variable $y=R_n/d$ and notations 
$t_1=k_1 d$ and $t_2=k_2 d$ we get 
\begin{equation}
\tilde{C}(k_1,k_2)=C(k_1 d,k_2 d),
\end{equation}
where
\begin{eqnarray}\label{Cuj}
C(t_1,t_2)&=&\frac{d^3}{a^3}\int\limits_1^{\infty} dy 
\frac{y^2-1}{y}\nonumber \\
&&\hspace{-1cm}\times \bigl [
C_{c+}(t_1,t_2,y) \cos[(t_1+t_2) y]\nonumber \\
&&\hspace{-0.8cm}+C_{c-}(t_1,t_2,y) \cos[(t_1-t_2) y]\nonumber \\
&&\hspace{-0.8cm}+C_{s+}(t_1,t_2,y) \sin[(t_1+t_2) y]\nonumber \\
&&\hspace{-0.8cm}+C_{s-}(t_1,t_2,y) \sin[(t_1-t_2) y]\bigr ].
\end{eqnarray}
The occurring integrals with respect to $y$ look like
\begin{subequations}
\begin{equation}
G_c(t,n):=\int\limits_1^{\infty} dy \frac{y^2-1}{y} 
\frac{\cos[t y]}{y^n},  
\end{equation}
and
\begin{equation}
G_s(t,n):=\int\limits_1^{\infty} dy \frac{y^2-1}{y} 
\frac{\sin[t y]}{y^n}.  
\end{equation}
\end{subequations}

Evaluating the $G_s$ and $G_c$ functions analytically by using MATHEMATICA, 
we get
\begin{eqnarray}\label{Cujabb}
&&\hspace{-1cm}C(t_1,t_2)=\frac{d^3}{a^3}\bigl [
f_{c+}(t_1,t_2) \cos[t_1+t_2]\nonumber \\
&&+f_{c-}(t_1,t_2) \cos[t_1-t_2]\nonumber \\
&&+f_{s+}(t_1,t_2) \sin[t_1+t_2]\nonumber \\
&&+(t_1-t_2) f_{s-}(t_1,t_2) \sin[t_1-t_2]\nonumber \\
&&\hspace{-0.5cm}+(t_1-t_2)^2 f_{ci}(t_1,t_2)
\bigl (Ci[t_1+t_2]-Ci[t_1-t_2]\bigr )\bigr ],
\end{eqnarray}
where for $t_1=t_2=t\gg 1$
\begin{eqnarray}
\label{fs}
  &&f_{c+}(t,t)\approx\frac{3825}{4 t^6}\nonumber \\
  &&f_{c-}(t,t)\approx -\frac{225}{2 t^4}\nonumber \\
  &&f_{s+}(t,t)\approx\frac{225}{2 t^5}\nonumber \\
  &&f_{s-}(t,t)\approx\frac{225}{2 t^4}\nonumber \\
  &&f_{ci}(t,t)\approx\frac{225}{2 t^4}
\end{eqnarray}
and $Ci[t]=-\int\limits_t^{\infty} du\frac{\cos u}{u}$ is the cosine
integral function.

Let us consider now the integration with respect to $R_{n'}$ in the first part
of Eq.~(\ref{Ksim}):
\begin{eqnarray}
\tilde{D}(k_3,k_4)&:=&\frac{1}{a^3}\int\limits_{r_0}^d dR_{n'} R_{n'}^2 
\nonumber \\
&&\hspace{-2.4cm}\times \bigl [3 v_0(k_4,R_{n'}) v_1(k_3,R_{n'}) + 
      3 v_0(k_3,R_{n'}) v_1(k_4,R_{n'}) \nonumber \\
     &&\hspace{-2.2cm}+ v_1(k_3,R_{n'}) v_1(k_4,R_{n'}) + 
      2 v_1(k_4,R_{n'}) v_2(k_3,R_{n'})\nonumber \\
     &&\hspace{-2.6cm}+ 2 v_1(k_3,R_{n'}) v_2(k_4,R_{n'}) + 
      4 v_2(k_3,R_{n'}) v_2(k_4,R_{n'})\bigr ].
\end{eqnarray}
After substituting the $v_m(k, R_{n'})$ matrix elements given by
Eq.~(\ref{vkm}) and introducing the dimensionless integration variable
$y'=R_{n'}/d$ and notations $t_3=k_3 d$, $t_4=k_4 d$, and $y_0=r_0/d$ we get
\begin{equation}
\tilde{D}(k_3,k_4)=D(k_3 d,k_4 d),
\end{equation}
where
\begin{eqnarray}\label{Duj}
D(t_3,t_4)&=&\frac{d^3}{a^3}\int\limits_{y_0}^1 dy' y'^2\nonumber \\
&&\hspace{-1cm}\times \bigl [
D_{c+}(t_3,t_4,y') \cos[(t_3+t_4) y']\nonumber \\
&&\hspace{-0.8cm}+D_{c-}(t_3,t_4,y') \cos[(t_3-t_4) y']\nonumber \\
&&\hspace{-0.8cm}+D_{s+}(t_3,t_4,y') \sin[(t_3+t_4) y']\nonumber \\
&&\hspace{-0.8cm}+D_{s-}(t_3,t_4,y') \sin[(t_3-t_4) y']\bigr ].
\end{eqnarray}
The occurring integrals with respect to $y'$ look like
\begin{subequations}
\begin{equation}
H_c(t,n):=\int\limits_{y_0}^1 dy' 
y'^2\frac{\cos[t y']}{y'^n}, 
\end{equation}
and
\begin{equation}
H_s(t,n):=\int\limits_{y_0}^1 dy' 
y'^2\frac{\sin[t y']}{y'^n}.  
\end{equation}
\end{subequations}

Evaluating the $H_s$ and $H_c$ functions analytically by using MATHEMATICA, 
we get
\begin{eqnarray}\label{Dujabb}
&&\hspace{-1cm}D(t_3,t_4)=\frac{d^3}{a^3}\bigl [
g_{c+}(t_3,t_4) \cos[t_3+t_4]\nonumber \\
&&+g_{c-}(t_3,t_4) \cos[t_3-t_4]\nonumber \\
&&+g_{s+}(t_3,t_4) \sin[t_3+t_4]\nonumber \\
&&+(t_3-t_4) g_{s-}(t_3,t_4) \sin[t_3-t_4]\nonumber \\
&&+h_{c+}(t_3,t_4,y_0) \cos[(t_3+t_4) y_0]\nonumber \\
&&+h_{c-}(t_3,t_4,y_0) \cos[(t_3-t_4) y_0]\nonumber \\
&&+h_{s+}(t_3,t_4,y_0) \sin[(t_3+t_4) y_0]\nonumber \\
&&+(t_3-t_4) h_{s-}(t_3,t_4,y_0) \sin[(t_3-t_4) y_0]
\bigr ],
\end{eqnarray}
where for $t_3=t_4=t\gg 1$
\begin{eqnarray}
  \label{gs}
 &&g_{c+}(t,t)\approx\frac{225}{2 t^4}\nonumber \\
 &&g_{c-}(t,t)\approx\frac{225}{2 t^4}\nonumber \\
 &&g_{s+}(t,t)\approx -\frac{1125}{t^5}\nonumber \\
 &&g_{s-}(t,t)\approx\frac{1125}{2 t^6}
\end{eqnarray}
and
\begin{subequations}
\label{hs}
\begin{eqnarray}
&&\hspace{-1.2cm}h_{c+}(t_3,t_4,y_0)=
\frac{20250}{t_3^5 t_4^5 y_0^7} - \frac{8100}{t_3^3 t_4^5 y_0^5} -
    \frac{20250}{t_3^4 t_4^4 y_0^5}\nonumber \\&&\hspace{1cm}
 - \frac{8100}{t_3^5 t_4^3 y_0^5} +\frac{
    1350}{t_3^2 t_4^4 y_0^3} + \frac{6525}{2 t_3^3 t_4^3 y_0^3} \nonumber
    \\&&\hspace{1cm}+
    \frac{1350}{t_3^4 t_4^2 y_0^3} - \frac{225}{2 t_3^2 t_4^2 y_0},
\end{eqnarray}
\begin{eqnarray}
&&\hspace{-1.2cm}h_{c-}(t_3,t_4,y_0)=
-\frac{20250}{t_3^5 t_4^5 y_0^7} + \frac{8100}{t_3^3 t_4^5 y_0^5} -
    \frac{20250}{t_3^4 t_4^4 y_0^5} \nonumber \\&&\hspace{1cm}
+ \frac{8100}{t_3^5 t_4^3 y_0^5} +
    \frac{1350}{t_3^2 t_4^4 y_0^3} - \frac{6525}{2 t_3^3 t_4^3 y_0^3}
    \nonumber \\&&\hspace{1cm}+
    \frac{1350}{t_3^4 t_4^2 y_0^3} - \frac{225}{2 t_3^2 t_4^2 y_0},
\end{eqnarray}
\begin{eqnarray}
&&\hspace{-1.2cm}h_{s+}(t_3,t_4,y_0)=
\frac{20250}{t_3^4 t_4^5 y_0^6} + \frac{20250}{t_3^5 t_4^4 y_0^6} -
    \frac{1350}{t_3^2 t_4^5 y_0^4} \nonumber \\&&\hspace{1cm}
- \frac{8100}{t_3^3 t_4^4 y_0^4} -
    \frac{8100}{t_3^4 t_4^3 y_0^4} - \frac{1350}{t_3^5 t_4^2 y_0^4}
    \nonumber \\&&\hspace{1cm}+
    \frac{1125}{2 t_3^2 t_4^3 y_0^2} + \frac{1125}{2 t_3^3 t_4^2 y_0^2},
\end{eqnarray}
and
\begin{eqnarray}
&&\hspace{-1.2cm}h_{s-}(t_3,t_4,y_0)=
-\frac{20250}{t_3^5 t_4^5 y_0^6} + \frac{1350}{t_3^3 t_4^5 y_0^4}
    - \frac{6750}{t_3^4 t_4^4 y_0^4} \nonumber \\&&\hspace{1cm}
+ \frac{1350}{t_3^5 t_4^3 y_0^4} -
    \frac{1125}{2 t_3^3 t_4^3 y_0^2}.
\end{eqnarray}
\end{subequations}


\begin{thebibliography}{}
\bibitem{Riemann} The formula can be proved by integration by part.
\bibitem{xnagy} For $x\gg 1$ the approximate expression Eq.~(\ref{Pkoz}) is
  not valid, for e.g. $x=10$ we get
  $\mathrm{P}^{\mathrm{num}}(x=10,\omega=0.01)=8992.5$ while
  $\mathrm{P}^{\mathrm{appr}}(x=10,\omega=0)=-1297.3$.
\end{thebibliography}
\end{document}